\documentclass[sn-mathphys-num]{sn-jnl}

\usepackage{graphicx}%
\usepackage{multirow}%
\usepackage{amsmath,amssymb,amsfonts}%
\usepackage{amsthm}%
\usepackage{mathrsfs}%
\usepackage[title]{appendix}%
\usepackage{xcolor}%
\usepackage{textcomp}%
\usepackage{manyfoot}%
\usepackage{booktabs}%
\usepackage{algorithm}%
\usepackage{algorithmicx}%
\usepackage{algpseudocode}%
\usepackage{listings}%

\newcommand{\e}{\mathrm{e}}
\renewcommand{\i}{\mathrm{i}}
\usepackage{xcolor, soul}
\definecolor{lightyellow}{rgb}{1, 1, 0.7}   
\sethlcolor{lightyellow}                    
\usepackage[textsize = small, backgroundcolor = lightyellow, linecolor = gray, disable]{todonotes}  %
\usepackage[export]{adjustbox}  
\usepackage{siunitx}
\usepackage{tabularx}   
\usepackage{subcaption}   
\usepackage{soul}  
\usepackage{braket}
\usepackage{xfrac}

\newcommand\hlbreakable[1]{\textcolor{red}{#1}}
\newcommand\hlnew[1]{\colorbox{lightyellow}{\textcolor{black}{#1}}}
\renewcommand\hlbreakable[1]{{#1}}
\renewcommand\hlnew[1]{{#1}}

\theoremstyle{thmstyleone}%
\theoremstyle{thmstyletwo}%

\theoremstyle{thmstylethree}%

\begin{document}

\title[Ramsey-Bord\'e  Atom Interferometry with a thermal strontium beam]{Ramsey-Bord\'e  Atom Interferometry with a thermal strontium beam for a compact optical clock}


\author*[1]{\fnm{Oliver} \sur{Fartmann}}\email{fartmann@physik.hu-berlin.de}

\author[1]{\fnm{Martin} \sur{Jutisz}}
\author[1]{\fnm{Amir} \sur{Mahdian}}
\author[1]{\fnm{Vladimir} \sur{Schkolnik}}
\author[1]{\fnm{Ingmari C.} \sur{Tietje}}
\author[1]{\fnm{Conrad} \sur{Zimmermann}}
\author[1,2]{\fnm{Markus} \sur{Krutzik}}

\affil[1]{\orgdiv{Institut für Physik}, \orgname{Humboldt-Universität zu Berlin}, \orgaddress{\street{Newtonstr. 15}, \city{Berlin}, \postcode{12489}, \state{Berlin}, \country{Germany}}}

\affil[2]{\orgname{Ferdinand-Braun-Institut (FBH)}, \orgaddress{\street{Gustav-Kirchhoff-Straße 4}, \city{Berlin}, \postcode{12489}, \state{Berlin}, \country{Germany}}}


\abstract{
Compact optical atomic clocks have become increasingly important in field applications and clock networks.
Systems based on Ramsey-Bordé interferometry (RBI) with a thermal atomic beam seem promising to fill a technology gap in optical atomic clocks, as they offer higher stability than optical vapour cell clocks while being less complex than cold atomic clocks. \\
Here, we demonstrate RBI with strontium atoms, utilizing the narrow ${}^1\!S_0 \rightarrow {}^3\!P_1$ intercombination line at \SI{689}{nm}, yielding a \SI{60}{kHz} broad spectral feature. The obtained Ramsey fringes for varying laser power are analyzed and compared with a numerical model. The ${}^1\!S_0 \rightarrow {}^1\!P_1$ transition at \SI{461}{nm} is used for fluorescence detection. Analyzing the slope of the RBI signal and the fluorescence detection noise yields an estimated short-term stability of $<\num{4e-14} / \sqrt{\tau / \SI{1}{s}}$. We present our experimental setup in detail, including the atomic beam source, frequency-modulation spectroscopy to lock the \SI{461}{nm} laser, laser power stabilization and the high-finesse cavity pre-stabilization of the \SI{689}{nm} laser. \\
Our system serves as a ground testbed for future clock systems in mobile and space applications.
}




\maketitle


\section{Introduction}
\label{sec:intro}
Due to their exceptional stability and accuracy, optical atomic clocks are a cornerstone of a broad range of applications including global navigation satellite systems (GNSS) \cite{Schuldt2021, Batori2021}, GNSS-denied navigation \cite{Sheerin2020}, data synchronisation \cite{Morzynski2016} and relativistic geodesy \cite{McGrew2018, Mehlstaubler2018, Beloy2021}. Additionally, optical atomic clocks are at the core of fundamental research as probes of variations in fundamental constants \cite{Dzuba2012, Lange2021}, ultra-light dark matter searches \cite{Derevianko2016, Beloy2021}, gravitational wave detection \cite{Kolkowitz2016}, tests of Einstein's Equivalence Principle \cite{Ahlers2022, Safronova2018a, Delva2017, Lange2021} and the redefinition of the SI units \cite{Lodewyck2019}. 
The performance of optical atomic clocks has surpassed microwave clocks due to their higher $Q$-factor defined as $Q = \nu/\Delta \nu$. With the development of the frequency comb, the stability of optical clocks could be transferred to the microwave regime which is accessible to electronic readout (see \cite{Fortier2019} for review). \\
The development of compact and portable optical atomic clocks suitable for deployment in harsh environments remains a major challenge as of yet. Vapour cell based references and clocks have already been demonstrated under harsh conditions such as sounding rockets and are currently developed towards satellite operation  \cite{Doringshoff2019, Kuschewski2024}.
Due to collisional broadening, the employed frequency discriminator of these systems is commonly several \SI{100}{kHz} broad. Rubidium two-photon clocks have demonstrated instabilities of $\sigma_y(\tau) = 4 \times 10^{-13} / \sqrt{\tau / \SI{1}{s}}$ averaging down to $\sigma_y(\tau) < 10^{-14}$ \cite{Martin2018a}. Iodine clocks have demonstrated instabilities $\sigma_y(\tau) < 10^{-14}$ between \SI{1}{s} and \SI{1000}{s} \cite{Schuldt2017}. \\
Lattice and ion clocks probe linewidths in the order of \SI{1}{Hz} due to the controlled atomic motion and reach instabilities of $\sigma_y(\tau) < 1 \times10^{-16}/\sqrt{\tau / \SI{1}{s}}$ averaging down to $\sigma_y(\tau) \sim 10^{-18}-10^{-19}$ \cite{Bothwell2019}. They are, however, still predominantly bound to controlled laboratory environments, although considerable progress is expected over the next decade \hlnew{\cite{Takamoto2020, Gellesch2020, Takamoto2022}}. Other approaches include superradiance both on thermal atomic beams~\cite{Fama2024, Tang2023} and cold atomic ensembles~\hlnew{\cite{Bohr2023, Okaba2019}}, two-photon spectroscopy on the calcium ${}^1\!S_0 \rightarrow \,{}^{1}\!D_{2}$ transition~\cite{Jackson2023} and three-photon excitation of the ${}^1\!S_0 \rightarrow \,{}^{3}\!P_{0}$ transition in strontium~\cite{Carman2024}. \\
Clocks based on the interrogation of atomic beams with Ramsey-Bordé interferometry (RBI) \cite{Borde1984a} present an intermediate solution in both complexity and performance. \\ 
RBI with calcium atoms has been demonstrated to offer instabilities of $\sigma_y(\tau) < 10^{-15}$ between \SI{1}{s} and \SI{1000}{s} with a continuous read-out \cite{Olson2019b, Hemingway2020a}.
Alkaline earth metals are a natural choice as they offer narrow intercombination transitions as well as sufficiently broad transitions for detection. 
In the past, next to calcium~\cite{Olson2019b, McFerran2010, Sheerin2020, Kersten1999, Ito1991}, magnesium~\cite{Sterr1992}, sulfur hexafluoride~\cite{Borde1984a} and Li$^{+}$-ions~\cite{Sun2023} have been employed for RBI in various groups. \\ 
Here, we demonstrate RBI with a thermal beam of strontium atoms.
Compared to the most commonly used element calcium, strontium has a broader natural linewidth for the clock transition ($\Gamma = 2\pi \,{\times}\, \SI{7.4}{kHz}$ vs. $2\pi \,{\times}\, \SI{375}{Hz}$) and a higher vapour pressure, which decreases the oven temperature by about \SI{150}{\celsius} \cite{Asano1978, Rudberg1934}. 
Furthermore, a higher atomic mass leads to a lower relativistic Doppler shift ($\sim \SI{430}{Hz}$ vs. $\sim \SI{940}{Hz}$). 
It might be a promising candidate for compact setups and could also be used as a pre-stabilization for strontium lattice clocks, which currently rely on high-performance cavity pre-stabilization of the clock laser. \\
\hlbreakable{The goal of the presented experiment is developing a compact clock for mobile and space applications. The focus of this paper is presenting the experimental setup and demonstrating Ramsey-Bordé fringes using strontium. Long-term stability measurements and characterizing systematic shifts to the clock frequency will be subject of future work, with a target instability of $\sigma_y(\tau) < 10^{-15}$ which has been demonstrated with a similar but larger system employing calcium in Ref. \cite{Olson2019b}.} \\
The paper is organized as follows. Section \ref{sec:system_overview} will provide a brief introduction to RBI and an overview of the experimental setup. 
Section \ref{sec:atomic_beam} presents the atomic beam source and its characterization. 
The spectroscopy chamber is presented in section \ref{sec:spectroscopy_chamber}.
Section \ref{sec:atom_interferometer} details the atom interferometer spectra and their numerical analysis.
In section \ref{sec_detection} the electron-shelving fluorescence detection is presented with an estimation of its influence on the clock's short-term stability. 
We conclude with a discussion of the results and an outlook in section \ref{sec:conclusion}. \\
Appendix \ref{sec:laser_stabilization} covers the power stabilization of the clock- and the detection laser, the pre-stabilization of the clock laser to a high-finesse cavity as well as frequency stabilization of the detection laser to the atomic transition via frequency-modulation spectroscopy. 
\begin{figure}[!t]
\begin{tabularx}{\columnwidth}[b]{Xr}

    \begin{subfigure}[b]{0.45\columnwidth}
    \includegraphics[width=\hsize, right]{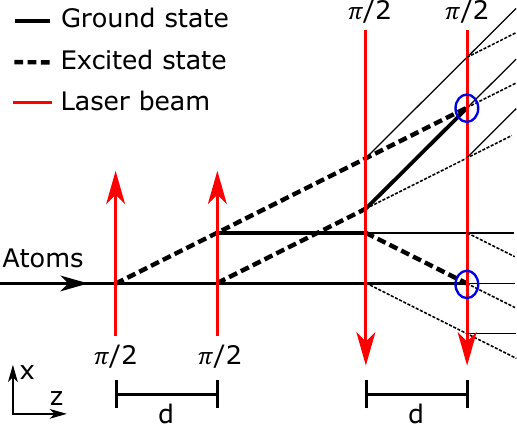} 
    \caption{}
    \label{fig_sketch_AI}
    \end{subfigure}
 &
    \begin{subfigure}[b]{0.45\columnwidth}
    \includegraphics[width = 0.9 \hsize, center]{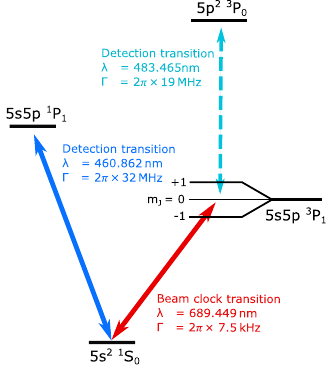}
    \caption{}  
    \label{fig:termscheme}
    \end{subfigure}     
\end{tabularx} 
\caption{
\textbf{(a)} Schematic of Ramsey-Bordé interferometry. Laser beams, indicated by red arrows, are perpendicular to the atomic beam. As the atoms traverse each laser beam, they experience a $\pi /2$ pulse, which splits the wavefunction. Recombination of the wavefunction leading to interference occurs at the last laser beam, indicated by the blue circles. The excited state probability after the interferometer has a sinusoidal dependency on the laser frequency detuning.  \\
\textbf{(b)} Term scheme of $^{88}$Sr. The $5s^2\,{}^1\!S_0 \rightarrow 5s5p \,^{3}\!P_{1}$ intercombination line at \SI{689}{nm} serves as the clock transition. For the detection through electron-shelving, there are two options - the  $5s^{2} \,^{1}\!S_{0} \rightarrow 5s5p{}\,^{1}\!P_{1}$ transition at \SI{461}{nm} which is employed here and the  $5s^{2}\,{}^{3}\!P_{1} \rightarrow 5p^{2}\,{}^{3}\!P_{0}$ transition at \SI{483}{nm}. 
}
\label{fig_setup}
\end{figure}


\section{System overview} 
\label{sec:system_overview}
\begin{figure*}[t]
    \centering
    \includegraphics[width = \linewidth]{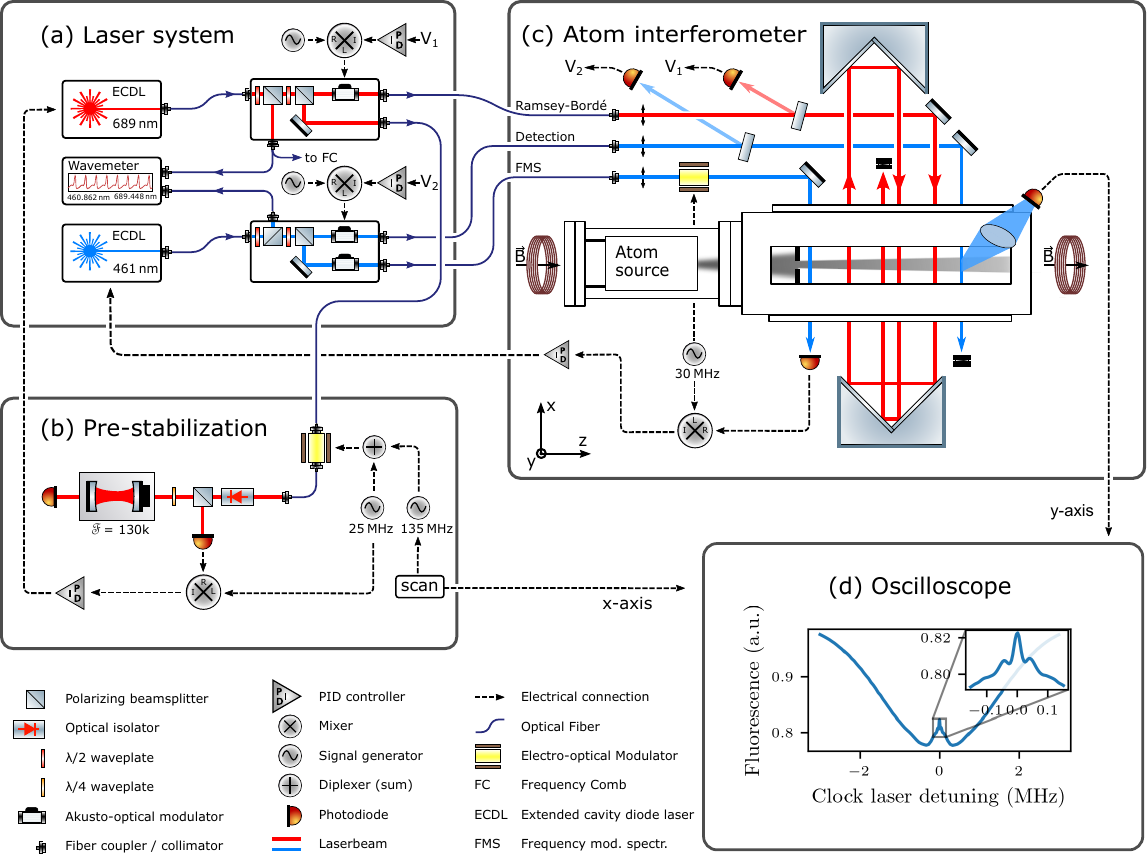} %
    \vspace{1mm}
    \caption{Schematic representation of the experimental setup.
    \textbf{(a)} ECDLs at \SI{689}{nm} and \SI{461}{nm} serve as clock and detection laser, respectively.
    \textbf{(b)} The clock laser is pre-stabilized to a high-finesse cavity by employing a dual-sideband lock with a variable offset frequency of around \SI{135}{MHz}. 
    \textbf{(c)} A fraction of the clock and detection laser beams are used for power stabilization. The remaining clock laser light is guided through the atomic beam by two hollow roof prisms, forming the atom interferometer as in figure \ref{fig_sketch_AI}. The detection laser frequency is locked to transition by frequency-modulation spectroscopy (FMS). Another laser beam passes the atomic beam after the interferometer where the fluorescence is monitored by a photodiode. 
    \textbf{(d)} Scanning the clock laser frequency via the offset frequency in the pre-stabilization and recording the fluorescence signal reveals optical Ramsey fringes on an oscilloscope.}
    \label{fig_overview}
\end{figure*}

\subsection{Ramsey-Bordé Interferometry (RBI)}
Here we give a brief introduction to RBI. A more detailed description can be found in \cite{Strathearn2022, Borde1984a, Sterr1997}. A schematic representation of RBI is depicted in figure \ref{fig_sketch_AI}. 
Each atom interacts coherently with four perpendicular laser beams with a wavelength of \SI{689}{nm}, where the second pair of beams is counter-propagating to the first pair. This can also be understood as two succeeding Ramsey interferometers. The interaction leads to a consecutive splitting and recombination of the atomic wavefunction, forming two closed atom interferometer ports, highlighted by blue circles in figure \ref{fig_sketch_AI}.
For a single atom and the clock laser frequency close to resonance, the excitation probability $P_e$ of both interferometer ports has a periodic dependence on the clock laser frequency detuning $\Delta \nu = \nu -\nu_0$ given by \cite{Olson2019b}
\begin{align}
P_e \sim \cos{\left[4 \pi \,(\Delta \nu \pm \delta \nu_\text{recoil} + \nu_\text{sys}) \, T  +\phi \right]} 
\label{eq:excitation_prob}
\end{align} 
with distance $d$  between two co-propagating laser beams, longitudinal velocity $v$, free evolution time $T=v/d$ and recoil frequency $\delta\nu_\text{recoil} = \tfrac{h}{2m \lambda^2} $. Other systematic frequency shifts including time dilation are included in $\nu_\text{sys}$. The global phase $\phi = \phi_4-\phi_3+\phi_2-\phi_1 $ accounts for the light phases $\phi_i$ at the four interaction zones, which are imprinted onto the atoms. Assuming the distances between both pairs of co-propagating laser beams are equal, the first-order Doppler shift for both Ramsey interferometers cancels and does not affect the fringe phase. \\
For a thermal ensemble of atoms, the axial velocity and thus the period $T$ varies. Therefore, the sinusoidal signals from different atoms only add up constructively around resonance. The slope of the fringes can serve as the frequency discriminator for the atomic clock. The FWHM linewidth of the fringes can be approximated from equation~\ref{eq:excitation_prob} to 
\begin{align} \label{eq_fringe_width}
\Delta \nu_\text{FWHM} \approx \frac{1}{4T} \,\text{.}
\end{align}
A large distance $d$ leads to exponential decay of the excited state and thus to a degradation of the signal amplitude. Meanwhile, the linewidth decreases anti-proportional to the distance $d$. Thus there exists a finite distance $d_\text{opt}$ which maximizes the steepness of the fringe slope. For strontium atoms with a mean velocity of \SI{420}{m/s} the optimum distance is $d_\text{opt} \approx \SI{3.6}{mm}$ leading to a linewidth of $\Delta \nu_\text{FWHM} = \SI{29}{kHz}$, based on Monte-Carlo simulations using formulas derived in \cite{Borde1984a}. This corresponds to around four times the natural transition linewidth of the clock transition $\Delta \nu_\text{nat} = 7.5$~kHz. 
Due to the interaction time broadening of the individual laser-atom interactions, the laser beams address a wide range of transversal velocity classes, leading to low atomic shot noise.\\ 
The population distribution of the involved electronic states is typically measured with a laser beam interrogating a broad transition connected to either the ground or excited clock state, commonly referred to as electron shelving detection \cite{Manai2019, Kai-Kai2006}  (see figure \ref{fig:termscheme}). This ensures that many photons are emitted per atom reducing both photonic shot noise and technical noise  of the fluorescence detection.

\subsection{Experimental setup}
The relevant level structure of strontium is depicted in figure \ref{fig:termscheme}. The   $^1S_0 \rightarrow \, ^3P_1$ intercombination line at \SI{689}{nm} is chosen for Ramsey-Bordé interferometry. To probe the population of the ground state behind the interferometer, we detect the fluorescence of the $^1S_0 \rightarrow \, ^1P_1$ transition, which is driven by a laser locked to the resonance at \SI{461}{nm}. 
A schematic representation of the experiment is shown in Figure \ref{fig_overview}. The clock and detection transitions are both addressed by extended cavity diode lasers (ECDLs). 
We use optical divider boards to split up the laser beams for frequency stabilization, wavemeter readings, and the spectroscopy setup. \\
The \SI{689}{nm} laser frequency is pre-stabilized to a high finesse cavity. By employing a dual-sideband lock (see section \ref{sec_cavity}), a tunable offset frequency of around \SI{135}{MHz} between the laser frequency and the resonance mode of the cavity is realized. 
The spectroscopy setup comprises a vacuum chamber maintained at a pressure of \SI{3e-6}{mbar}. Inside the vacuum chamber, an in-house atomic oven in conjunction with an aperture creates a strontium atomic beam with an axial intensity of \SI{1.4e+14}{s^{-1} sr^{-1}} (see section \ref{sec:atomic_beam} for details).
The \SI{689}{nm} laser beam traverses the atomic beam and is retroreflected twice by hollow roof prisms. 
Two coils generate a magnetic field of approximately $B \approx \SI{8}{Gs}$ along the atomic beam to split the $^3P_1$ state into three Zeeman sub-levels. All laser beams in the spectroscopy setup are p-polarized such that only the magnetically insensitive $\ket{J = 0, m_J=0} \rightarrow \ket{J' = 1, m_J' = 0}$ transitions are addressed. 
The resulting population of the ground state after the atom interferometer is measured via fluorescence at \SI{461}{nm}. Scanning the offset frequency of the \SI{689}{nm} laser over resonance depletes the ground state population and thus reveals Ramsey-Bordé fringes (see figure \ref{fig:signal}).


\section{Atomic beam source}
\label{sec:atomic_beam}
\begin{figure}[!t]
\begin{tabularx}{\columnwidth}{XXr}
\begin{tabular}[b]{l}
    \begin{subfigure}[b]{0.4\columnwidth}
    \includegraphics[width=0.95\linewidth, center]{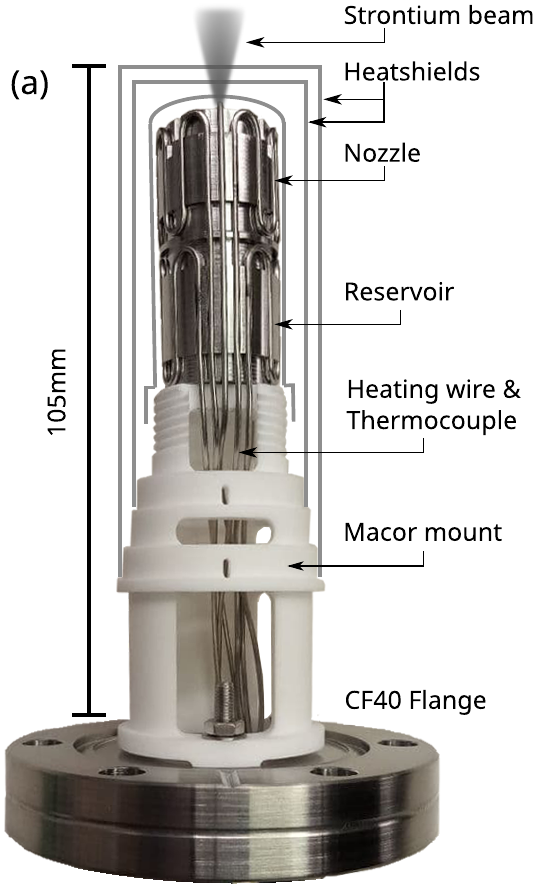}
    \vspace{-0.05cm}
    \end{subfigure}  
\end{tabular}
\hfill
&
\begin{tabular}[b]{l}
    \begin{subfigure}[b]{0.45\columnwidth}
    \includegraphics[ center]{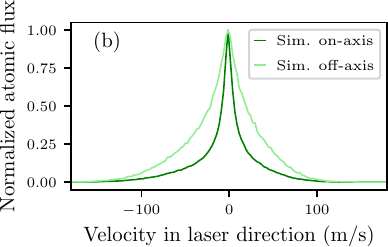} 
    \vspace{0.1cm}
    \end{subfigure} \\
    \begin{subfigure}[b]{0.45\columnwidth}
    \includegraphics[center]{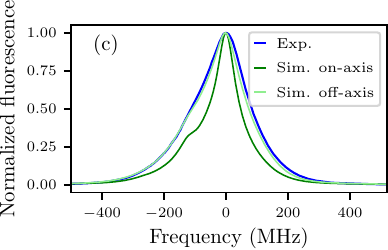} 
    \end{subfigure}
\end{tabular}

\end{tabularx}
\caption{
\textbf{(a)} Compact strontium oven. The reservoir (nozzle) is temperature stabilized to \SI{430}{\celsius} (\SI{510}{\celsius}) by a vacuum-compatible mineral-insulated heating wire. A mount made from Macor ceramics and two heat shields reduce the heat load to $<\SI{10}{W}$. \\
\textbf{(b)} Simulation of the atomic velocity distribution in laser direction. On-axis: the laser beam perfectly intersects the atomic axis. Off-axis: the laser beam is displaced w.r.t. the atomic axis by \SI{4}{mm} in the vertical direction. \\
\textbf{(c)} Fluorescence measurement of the atomic beam in an oven test chamber. The experimental data fit the theoretical expectation assuming the laser beam was displaced. See text for more details. 
}
\label{fig_oven_complete}
\end{figure}

\noindent
A key requirement for the RBI scheme is a collimated stream of strontium atoms. We have developed an atomic oven to generate the atomic beam, shown in figure \ref{fig_oven_complete}. While the mount between the vacuum flange and the oven assembly is made from Macor ceramics, all other components are manufactured from stainless steel. The reservoir contains \SI{2.5}{g} dendritic strontium and is heated to \SI{430}{\celsius} by vacuum-compatible mineral-insulated heating wire
meandering around the reservoir. 
An array of capillaries with a length of \SI{17}{mm}, inner diameter of \SI{200}{\micro\metre} and outer diameter of \SI{300}{\micro\metre} 
is used to collimate the atomic beam. The array is formed by clamping the capillaries into a \SI{60}{\degree} cut to ensure a close packing \cite{Senaratne2015}. An aperture of \(1{\times}\SI{5}{mm^2}\) (W\({\times}\)H) between the reservoir and the capillaries is used to select a part of the atomic beam. The capillaries are heated to \SI{510}{\celsius} with a separate heating wire. 
The temperature is measured at the bottom of the reservoir and the top of the nozzle with K-type mineral-insulated thermocouples. To reduce heat load, the reservoir is mounted onto a piece of Macor ceramics utilizing its low thermal conductivity of \SI{1.46}{W\, m^{-1}\,K^{-1}}. Two polished heat shields around the oven assembly reduce the heat loss from thermal radiation. The temperature stabilization limits residual fluctuations to $\SI{<10}{mK}$. The power consumption is $\SI{<10}{W}$. The oven fits into a standard CF40 tube with an outer diameter of the flange of \SI{70}{mm} and a length of \SI{126}{mm}.\\
The atomic flux was characterized in a dedicated vacuum system. A laser beam at \SI{461}{nm} with $d_{1/e^2} = \SI{2}{mm}$ diameter and $P = \SI{1.26}{mW}$ laser power intersects the atomic beam in a distance of \SI{103}{mm} from the oven aperture. The fluorescence is imaged onto a CCD camera while scanning the laser frequency. From this the phase-space density can be calculated as described in \cite{Jutisz2021}, similar to the procedure in \cite{Wouters2016}. Figure \ref{fig_oven_complete}(c) shows the integrated fluorescence signal plotted against the laser frequency detuning. \\
From the theoretical emission of a capillary \cite{Beijerinck1975} 
the expected velocity distribution in laser beam direction is calculated as detailed in Appendix \ref{app_oven}, see figure \ref{fig_oven_complete}(b). 
The expected velocity distributions are convoluted with the power-broadened natural linewidth which yields the expected fluorescence signal as shown in figure \ref{fig_oven_complete}(c).
We find the expected fluorescence signal matches well with the experimentally obtained values if we assume that the laser beam is displaced w.r.t. the atomic axis by \SI{4}{mm} in the vertical direction. \\
The cone angle of the atomic beam is given by $\theta_\text{FWHM} = \SI{20}{mrad}$ and the velocity distribution in laser direction has a width of $\Delta v_\text{x, FWHM} = \SI{14}{m/s}$. 
The oven is calculated to last for \SI{10}{years} before depleting.
The flux of atoms interacting with the laser beam was measured by integrating the fluorescence over the laser frequency detuning while taking into account the CCD quantum efficiency, laser power and photon collection efficiency. Using a Monte-Carlo simulation the fraction of atoms exiting the oven that interact with the laser beam is estimated, leading to an estimated total atomic flux of $\dot{N} = (1.5^{+3.0}_{-1.0}) \times 10^{14} \,\si{atoms \, s^{-1}}$. 
This yields an axial intensity of \(\displaystyle I(0) = \frac{\dot{N}}{\pi W} = (4.0^{+12.0}_{-2.7}) \times 10^{15} \,\si{atoms\, s^{-1} \,sr^{-1}}\), with the Clausing factor $W$.
For comparison, the axial intensity was independently measured in the Ramsey-Bordé setup via the absorption in the FMS path which yielded $(1.4^{+1.4}_{-0.7}) \times 10^{14} \,\si{atoms \,s^{-1}\, sr^{-1}}$. We suspect the latter to be lower due to slight misalignment of the atomic beam apertures. 


\section{Spectroscopy chamber} \label{sec:spectroscopy_chamber}
The spectroscopy chamber is made from titanium grade 5, which ensures low outgassing, low magnetic susceptibility and high specific strength.
The atomic beam is spatially filtered by a $\num{1}{\times}\SI{5}{mm^2}$ (W${\times}$H) aperture placed at a distance of \SI{150}{mm} from the oven to reduce the number of atoms with a high transversal velocity which would only contribute to the background signal. 
\noindent
Windows with an open aperture of $\num{55}{\times}\SI{7}{mm^2}$ (L${\times}$H) are indium sealed on both sides for the laser beams to pass through, and on top for fluorescence collection. A \SI{3}{l/s} titanium ion getter pump maintains a pressure of \SI{3e-6}{mbar}. 
The spectroscopy setup is mounted on a $\SI{45}{cm}\times\SI{45}{cm}$ breadboard to damp vibrations and enclosed with a black hardboard to reduce the influence of external light and air turbulence, which leads to phase fluctuations and laser beam pointing errors.\\
The \SI{689}{nm} laser beam is collimated with $1/e^2$ diameter of \SI{2.35}{mm} vertically and between \SI{0.96}{mm} and \SI{1.07}{mm} horizontally for the four interaction regions. Two one-inch hollow roof prisms are used as retroreflectors, setting the Ramsey beam distance to $d_{RI} = \SI{2.5\pm 0.5}{mm}$ and the central beam distance to $d_c = \SI{3.5\pm0.5}{mm}$. 
Hollow roof prisms enable a more compact beam path than a cat-eye configuration as the latter requires large focal lengths to avoid aberrations. One can show that the optical path phase term $\phi$ in equation \ref{eq:excitation_prob} is insensitive to alignment changes of the retroreflector as
\begin{align}
\phi = \phi_4-\phi_3+\phi_2-\phi_1 = 2 \pi \nu/c (L_{14} - L_{23}) + \phi_{wf}
\end{align}
with the optical path length $L_{14}$ ($L_{23}$) between the first (second) and fourth (third) laser-atom interaction. The term $\phi_{wf}$ accounts for a non-flat wavefront. Geometrically it is easy to see that the tilt and displacement of the prism change $L_{14}$ and $L_{23}$ by the same amount, leaving $\phi$ invariant. \\
The employed prisms have an angle accuracy of \SI{<3}{\arcsecond}, a nominal wavefront distortion of $<\lambda \,@\,\SI{689}{nm}$ and a reflectance of $R>\SI{99.5}{\%}$. The wavefront distortion after passing the retroreflector and the vacuum windows was measured using a Shack-Hartmann sensor. Compared to an ideal Gaussian wavefront, the peak-valley value is $\text{PV} \approx \lambda/10$ and the mean deviation is $\text{RMS} \approx \lambda/30$, both measured over the $1/e^2$ diameter.


\section{Atom interferometer} \label{sec:atom_interferometer}

\begin{figure}[t!]
    \centering
    \includegraphics[]{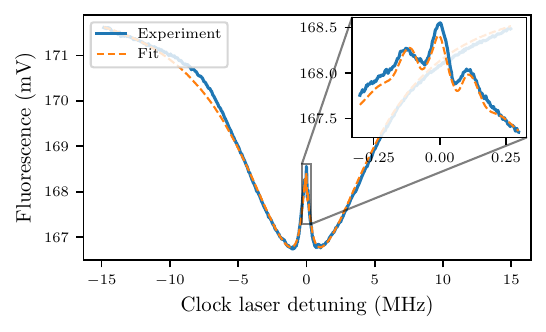} 
    \caption{Ramsey-Bordé fringes for a clock laser power of \SI{0.75}{mW}. Scanning the clock laser frequency and measuring the ground state population via the fluorescence of the \SI{461}{nm} transition reveals a Doppler-free signal with Ramsey fringes in the centre. Fitting the signal with equation \ref{eq_fit} yields a FWHM linewidth of \SI{60}{kHz} and a slope of \SI{15}{mV/MHz}.}
    \label{fig:signal}
\end{figure}

A typical Ramsey-Bordé interferometry measurement is shown in figure \ref{fig:signal}. For a better signal-to-noise ratio, ten traces were averaged. Due to the counter-propagating laser beams, a lamb-dip is visible in the centre of the Doppler broadened feature. Within the lamb dip, the sinusoidal Ramsey fringes are visible. \\
A numerical model is developed in appendix \ref{app_RB} to provide quantitative measures of the signal, which can be used to analyze the influence of different experimental parameters on the signal. We use these characterizations to maximize the fringe slope, which is the primary figure of merit for a frequency discriminator in an atomic clock. 
This model is briefly presented in the following. As discussed in appendix \ref{app_RB}, the fluorescence signal $F$ can be described by 
\begin{align} \label{equation_fit}
    F = V\,n_\gamma  \left( 1- \left(1+f_c \right)\, k\,f_b \right)
\end{align}
with a background function $f_b$ describing the incoherent envelope and a contrast function $f_c$ describing the fringes. $n_\gamma$ is the number of photons collected on the photodiode per atom and $V$ is the voltage signal per photon. The dimensionless factor $k$ accounts for the decay of the excited state. \\ 
\noindent
As further discussed in appendix \ref{app_RB} one can find the following approximations for the background function $f_b$ and contrast function $f_c$
\begin{align}
    f_b = 1 - \int_{-\infty}^{\infty} dv_t \,\frac{N(v_t)}{2} \left( 1+ \frac{1}{2sL(\nu-v_t/\lambda) + 2sL(\nu+v_t/\lambda) +1} \right)
\end{align}
with a Lorentzian transversal velocity distribution $N(v_t)$ and a Lorentzian lineshape $L(\nu\pm v_t/\lambda; \Delta \nu_h)$ for the homogeneous broadening $\Delta \nu_h$, which comprises the natural linewidth and interaction time broadening. The saturation parameter $s$ is the quotient of the laser-induced state transition rate on resonance to the  excited state decay rate, which includes natural decay and diffusion. This implies a laser power dependency of $s = \sqrt{P/P_\text{sat}}$ with saturation power $P_\text{sat}$ \\
The contrast function is modelled by
\begin{align} \label{eq:contrast_function}
    f_c  = \frac{\eta}{4} \int_{0}^{\infty} dv \,P_{v_L, \text{eff}}(v_L) \, \left(\cos(\Phi^+) + \cos(\Phi^-)\right) 
\end{align}
with a contrast quality factor $\eta$ and fringe phase $\Phi^\pm$. The effective longitudinal velocity distribution $P_{v_L, \text{eff}}(v_L)$ accounts for the received pulse area and decay of the excited state and is given by
\begin{align}
    P_{v_L, \text{eff}}(v_L) = P_{v_L}(v_L) \, \sin^4\left(A \frac{v_m}{v}\right) \,\exp\left(-\frac{D}{v} \Gamma\right)
\end{align}
with the pulse area $A$ for the mean velocity $v_m$ and the distance $D$ from the first clock laser interaction to the detection zone. The longitudinal velocity distribution $P_{v_l}$ is given by a standard Maxwell-Boltzmann distribution. 
The pulse area $A$ is defined as the polar angle of rotation of the Bloch vector \cite{Strathearn2022}.
For a resonant laser beam with a flat intensity profile, the pulse area is given by $A = 2\pi \Omega \tau = \pi \sqrt{P/P_\pi}$, with Rabi frequency $\Omega$, interaction time $\tau$ and laser power $P_\pi$ to drive a $\pi$-pulse. The influence of the laser frequency detuning on the pulse area is neglected. \\
The fringe phase $\Phi^\pm$ is given by 
\begin{align}
    \Phi^\pm = 4\pi \, \frac{d}{v} \left(\Delta\nu \pm \delta\nu_\text{recoil} + \frac{1}{2} \nu_0 \frac{v^2}{c^2}\right) + \phi .
\end{align}

\subsection{Experimental results}

\begin{figure}[t!]
\begin{tabularx}{\columnwidth}{XX}
    \begin{subfigure}[b]{0.53\columnwidth}
    \hspace*{-0.6cm}
    \includegraphics[]{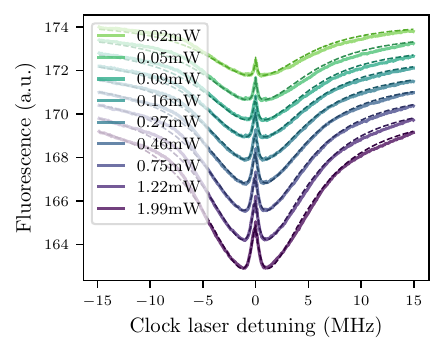} 
    \end{subfigure}  
&
    \begin{subfigure}[b]{0.37\columnwidth}
    \hspace*{0.4cm}
    \includegraphics[]{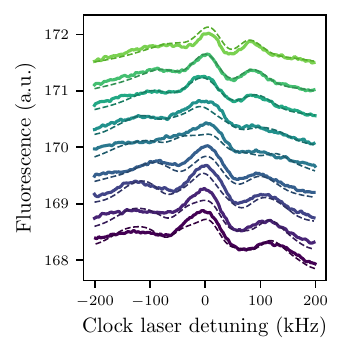}
    \end{subfigure} 
\end{tabularx}

    \begin{subfigure}[b]{0.45\columnwidth}
    \hspace*{-0.6cm}
    \includegraphics[]{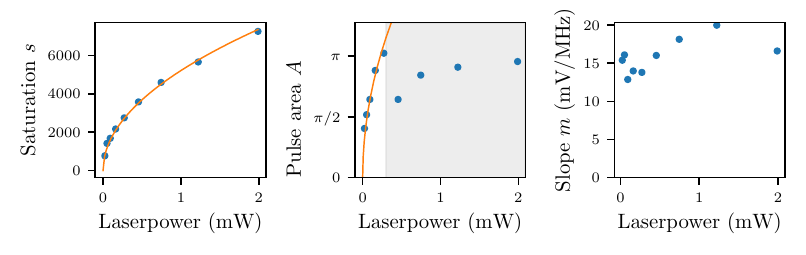}
    \end{subfigure} 

\caption{\textbf{Upper left:} Ramsey-Bordé signal for varying laser powers. The signals are shifted vertically w.r.t. each other for better visibility. Solid lines are experimental data, dashed lines are a theoretical fit with equation \ref{equation_fit}. \textbf{Upper right:} Close-up plot around the resonance frequency, Ramsey fringes are visible. \textbf{Lower:} Extracted parameters from the fit are plotted against the laser power. See text for details. }
\label{fig_laserpower}
\end{figure}

We obtained spectra for varying clock laser power from \SI{23}{\micro \W} to \SI{1.99}{\milli\W}, see figure \ref{fig_laserpower}. Spectra for lower laser power were not obtainable as the power stabilization becomes unstable below \SI{23}{\micro \W}. With increasing laser power, the signal becomes larger and the lamb dip widens due to saturation broadening. The Doppler broadening remains constant as it is not dependent on laser power. \\
The Ramsey fringe period increases with laser power. This is due to the fact that for increasing laser power and therefore increasing pulse area, faster atoms contribute to the fringe signal. Since these faster atoms have a shorter free precession time $T$, the fringe period increases. \\
For low pulse areas, the effective longitudinal velocity distribution $ \displaystyle P_{v_L, \text{eff}}(v_L)$  has a dominant contribution leading to a rather well-defined fringe period. As the laser power increases, the contribution becomes bi-modal where a slow velocity class receives a $\tfrac{3}{2}\pi$-pulse while a fast velocity class receives a $\frac{1}{2}\pi$-pulse. The signals from the slow velocity class with a small fringe period overlap with the signals from the fast velocity class with a large fringe period, leading to a deformation of the fringe signal. In this experiment, this can be observed for laser powers around \SI{150}{\micro \W}. \\
For a quantitative analysis, equation \ref{equation_fit} was fit to the spectra. Several parameter pairs have strong correlations, leading to an ambiguous fit. To prevent this, several parameters were fixed, see table \ref{tab:fit_parameter}. \\
The Doppler broadening width (FWHM) is between \SI{10.7}{MHz} and \SI{11.8}{MHz} which corresponds to between \SI{7.4}{m/s} and \SI{8.1}{m/s}. The atomic beam distribution calculation as described in \ref{app_oven} yields an expected width of \SI{4.4}{MHz} when the laser beam is perfectly perpendicular to the atomic beam. For a slight yaw angle the broadening increases by about \SI{5}{MHz/1\degree}. The yaw angle was optimized by centring the lamb dip with an estimated uncertainty of \SI{0.1}{\degree}.
\hlbreakable{The broadening and slight asymmetry of the transversal velocity distribution can be partially attributed to the atoms receiving recoil kicks from the FMS laser beam. This was tested by aquiring Ramsey-Bordé spectra for varying FMS laser power. The effect could be avoided by placing the FMS path downstream of the atom interferometer. However, no significant influence on the fringe signal was observed.}
\\
The saturation parameter $S$ and pulse area $A$ are plotted against the laser power in figure \ref{fig:signal}. The saturation follows $S = \sqrt{P/P_\text{sat}}$ with $P_\text{sat} = \SI{37}{\nano \W}$. The saturation intensity for a flat laser beam without diffusion is given by \(\displaystyle I_\text{sat} = \frac{\pi h c \Gamma}{3 \lambda^3}\) = \SI{3.0}{mW \per \centi \metre^2} \cite[p.142]{Foot2005}. Approximating the Gaussian laser beam as a flat beam with $1/e^2$ diameter yields an estimated saturation power of $P_\text{sat, exp} \approx \SI{68}{nW}$ which is comparable to the measured value. \\ 
The Ramsey fringes are well represented by the fit for lower laser power and the extracted pulse area follows $A = \pi \sqrt{P/P_{\pi}}$, where $P_{\pi} = \SI{230}{\micro\W}$.
With \( \displaystyle I = \frac{2 I_{sat}}{\Gamma^2} \Omega^2\), the laser power necessary to drive a $\pi$-pulse can be estimated to $P_{\pi} = \SI{110}{\micro \W}$ which is comparable to the measured value. \\
For higher laser powers, a discrepancy between the fit and the observed Ramsey fringes arises and the extracted pulse areas become unreliable. This is due to the simplification in equation \ref{eq:contrast_function} that the pulse area is independent of the laser frequency detuning. One can let \( \displaystyle A \rightarrow A \exp(-\tfrac{1}{4} ((2\pi \Delta \pm kv) \nu \, \tau_0)^2)\) as in \cite{Strathearn2022} to account for this. However, this requires introducing the transversal velocity distribution to the contrast function, which increases computation time which is not feasible for fitting. \\
The fringe width (FWHM) is about \SI{60}{kHz} which is close to the approximated fringe width of \SI{52}{kHz} as calculated from equation \ref{eq_fringe_width}.  \\
Lastly, the maximum slope $m$ of the Ramsey fringe is plotted against laser power in figure \ref{fig_laserpower}. A maximum of $m = \SI{20}{\milli \V \per \mega \Hz}$ is observed for \SI{1.22}{mW}.

\begin{table}[tbp] \centering
    \caption{Fixed fit parameter, see text for details. \vspace{0.5em}} 
    \begin{tabular}{@{}lll@{}}
    \toprule
    Parameter   & Value  &  Correlated to\\
    \midrule
    Oven temperature & \SI{450}{\celsius} & Ramsey distance \\
    Homogenous linewidth   &  \SI{7.5}{kHz} & Saturation parameter $s$ \\
    Ramsey distance $d$ & \SI{2.5}{mm} & Pulse area $A$ \\
    Distance between first interaction and detection zone $D$ & \SI{11}{mm} & Fringe contrast $\eta$ \\
    \bottomrule
    \end{tabular}
    \label{tab:fit_parameter}
\end{table}


\section{Detection} \label{sec_detection}
\begin{figure}
    \centering
    \includegraphics[]{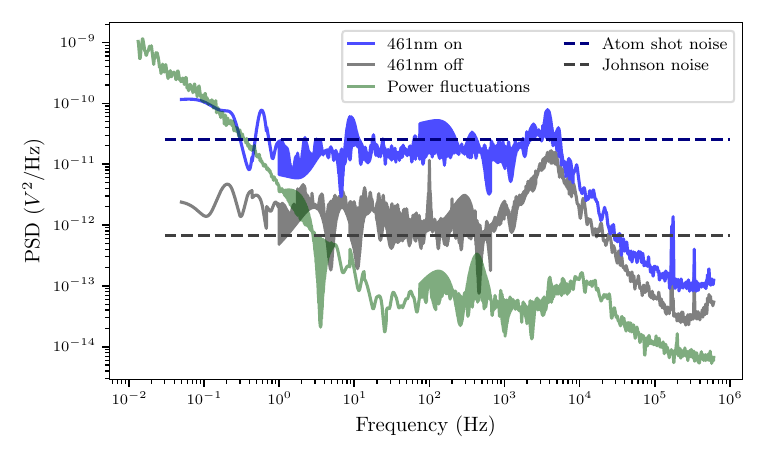}
    \caption{Power spectral density of the fluorescence detection noise. \hlbreakable{The blue and grey curve show the measured noise of the detector. The green curve is calculated from an out-of-loop measurement of the power stabilization described in the appendix \ref{powerstab}.
    For frequencies above \SI{1}{Hz}}, atomic shot-noise dominates and would limit the short-term stability in clock operation to $\sigma_y(\tau) \approx \num{4e-14}/\sqrt{\tau/\SI{1}{s}}$. See text for details.}
    \label{fig:fluor_noise}
\end{figure}

If the fringe signal with slope $m$ is used to lock the laser frequency to the atomic transition, residual fluorescence detection noise $\text{PSD}_{V}$ translates to residual frequency fluctuations of the laser frequency $\text{PSD}_{\nu}$
\begin{align} \label{eq_psd}
    \text{PSD}_{\nu} \geq \frac{\text{PSD}_{V}}{m^2} \, \text{.}
\end{align}
This is typically the dominant contribution to the short-term instability of the clock. Thus the fluorescence detection noise has to be minimized, with the fundamental limit given by atomic shot noise.  As mentioned before, the electron-shelving technique was utilized employing a detection beam at \SI{461}{nm} to probe the ground state. \\
The detection beam traverses the atomic beam \SI{3}{mm} after the clock laser interaction and has a $1/e^2$ diameter of $d_\text{det} = \SI{1.2}{mm}$. For strong saturation, every atom emits \( \displaystyle N_\gamma = \tfrac{1}{2} \frac{d_\text{det}}{\bar{v}} \Gamma \approx 220\) photons. \hlbreakable{Due to the $\pi$-polarization of the laser beam, fluorescence towards the atom interferometer is strongly suppressed \cite{Schioppo2012}.}
About $C_\text{eff} = \SI{2.5}{\%}$ of the fluorescence photons are collected on a photodiode, 
leading to the photonic shotnoise being negligible compared to the atomic shotnoise. \\
The photodiode detector design is similar to \cite{Boddy2014, Guttridge2016} with a trans-impedance of $\SI{40}{M\Omega}$. 
An inner floating ground reduces noise pick-off from ground loops and stray fields. The output of the trans-impedance amplifier and the inner ground signal are subtracted with a voltage pre-amplifier. The technical noise of the detector was measured and is comparable to the fundamentally limiting Johnson–Nyquist noise of the trans-impedance $\text{PSD}_{R} = 4 k_B T R = \SI{6.5e-13}{V^2/Hz} $, see figure \ref{fig:fluor_noise}. \\
For frequencies above \SI{1}{Hz}, the fluorescence detection noise is comparable to the atomic shot noise which is estimated by \(\displaystyle \text{PSD}_{\text{atom}} = 2 \frac{U^2 C_\text{eff} N_\gamma}{I_\text{PD}} = \SI{2.6e-11}{V^2/Hz} \), see figure \ref{fig:fluor_noise}. \\
For lower frequencies, the detection noise is dominated by residual \hlbreakable{drifts} of the detection laser power, which were measured in \ref{powerstab} \hlbreakable{and are attibuted to thermal effects on the photodetector used for power stabilization}. Residual frequency fluctuations of the detection laser calculated in section \ref{sec_FMS} are not a limiting factor. \\
Equation \ref{eq_psd} yields \( \displaystyle \text{PSD}_{\nu, \text{atom}} = \frac{\text{PSD}_{V, \text{atom}}}{m^2} = \SI{6e2}{Hz^2/Hz}\) which corresponds to an Allan deviation \cite[p.59]{Riehle2004} of 
\begin{align}
    \sigma_y(\tau) \approx \num{4e-14}/\sqrt{\tau/\SI{1}{s}}\,.
\end{align}
\hlbreakable{For longer time scales, slow drifts in the fluorescence signal can limit the clock's performance. To reduce sensitivity to these slow variations, lock-in detection is commonly employed \cite{McFerran2010, Olson2019b}. The methodology is identical to the FMS technique outlined in Appendix \ref{sec_FMS}. By using this method, the detection frequency - and thus the sensitivity to noise - is shifted to the modulation frequency. This can be set to around \SI{1}{kHz}, where noise is dominated by atomic shot noise. }

\section{Conclusion and outlook}
\label{sec:conclusion}
\noindent
This work demonstrates Ramsey-Bordé interferometry with a thermal strontium atomic beam. In this paper, we presented our system, including the compact atomic oven, the pre-stabilization of the clock laser to the cavity, the laser power stabilization as well as the frequency stabilization of the detection laser via frequency-modulation spectroscopy. The atom interferometer signal for varying laser powers was analyzed quantitatively and compared to numerical calculations. The fluorescence detection noise and its influence on the clock stability were characterized, which yields an estimated short-term stability of $\sigma_y(\tau) \approx \num{4e-14}/\sqrt{\tau/1s}$.  \\
The next step to realize an optical clock is to lock the laser to the fringes. To reduce the impact of fluorescence detection noise on long time scales, the error signal will be generated by modulating the clock laser frequency with an AOM and demodulating the fluorescence signal at \hlbreakable{the modulation frequency or its third harmonic}. A stability measurement is planned by comparing the Ramsey-Bordé clock against two independent two-photon Rb clocks at our laboratories. These measurements allow the study of systematic effects e.g. of the clock laser power, temperature and magnetic field strength on the clock frequency.\\
\hlbreakable{To decrease the relative atomic shot noise and therefore the short-term instability further, there are several options with varying degree of added complexity. The most straight forward is optimizing the collimation of the atomic beam, either by improving the atomic oven or by transversal laser cooling. A new oven design is being assembled aiming at higher mechanical robustness to survive harsh environments such as rocket starts as well as more flexibility regarding the collimation and refilling of the oven for test purposes \cite{Tietje2024}. 
Transversal cooling increases the number of atoms resonant with the clock laser by about a factor of 20 when using \SI{100}{mW} of cooling laser light over a distance of \SI{3}{cm}. Simultaneous transverse and longitudinal cooling as employed in Ref. \cite{Tobias2024} is not feasible for a strontium Ramsey-Bordé clock, as a reduction of the Ramsey distance due to reduced longitudinal velocity will lead to an overlap of the laser beams. \\
As was seen in figure \ref{fig:signal}, there is a large background signal stemming from atoms resonant to the detection beam at \SI{461}{nm} but not resonant to the clock laser beam at \SI{689}{nm} which contribute to the atomic shotnoise. Fluorescence detection of the excited state with a broad transition at \SI{483}{nm} can be utilized to reduce the number of background atoms. \\
Lastly, multifrequency spectroscopy as demonstrated recently in Ref. \cite{Tobias2024} can be utilized to address a larger fraction of the transversal velocity distribution. This method seems very promising as it is relatively simple to implement and increases the number of participating atoms significantly. We note that when the frequency spread of the modulated clock laser spans the full width of the transversal velocity distribution resonant to the detection laser, detection of the excited state population at \SI{483}{nm} does not have an advantage over detecting the ground state population at \SI{461}{nm}.
With implementing an improved atomic oven and multifrequency spectroscopy, we expect to achieve a short-term instability of $\sigma_y(\tau) < \num{1e-14}/\sqrt{\tau/\SI{1}{s}}$, possibly as low as $\num{1e-15}/\sqrt{\tau/\SI{1}{s}}$.}
For longer timescales, the dominant factor is expected to stem from optical phase fluctuations arising due to temperature fluctuations and air pressure fluctuations. These effects can be suppressed by placing the critical optical components within a temperature-stabilized vacuum system, for which we expect a long-term instability of $\sigma_y(\tau) < \num{1e-15}$.
Currently, such a system is being developed with a 19-inch form factor which will serve as a mobile demonstrator for network synchronisation. Simultaneously, another system is developed for future demonstration in space towards GNSS applications. \hlbreakable{Due to the continuous feedback and low short-term instability of the atomic signal, a less stable but more compact pre-stabilization of the clock laser can be utilized such as a monolithic cavity \cite{Zhang2024}.} \\


\clearpage
\begin{appendices}

\section{Laser stabilization} \label{sec:laser_stabilization}

\subsection{Power stabilization} \label{powerstab}
The clock laser power is stabilized to ensure a consistent pulse area. 
The detection laser power is also stabilized as power fluctuations translate to fluorescence signal noise (see figure \ref{fig:fluor_noise}).
The setup for power stabilization is identical for both lasers, albeit a \SI{1.5}{MHz} frequency offset of the acousto-optic modulators (AOMs) to reduce crosstalk. After polarization cleaning, a beam sampler reflects 3\% of the optical power onto a photodiode.
The beam sampler is placed with a low angle between its normal and the laser beam to minimize sensitivity to residual polarization fluctuations. A \SI{5}{cm} long lens tube together with a \SI{12.7}{mm} wide aperture is used to reduce the ambient light that reaches the photodiode. The photodiode signal is fed into a PI$^2$ controller with integrator corner frequencies of $f_{I_1} = \SI{100}{kHz}$ and $f_{I_2} = \SI{100}{Hz}$. The control signal is then fed into the I-port of a mixer. \\
An \SI{80}{MHz} RF signal is fed into the R-port of the mixer
(see figure \ref{fig_overview}). The DC voltage from the PI$^2$ controller, applied to the I-port, determines the RF amplitude level of the output at the L-port. This output is fed into the acousto-optical modulators and determines the light power guided to the spectroscopy setup. \\ 
The residual power fluctuations were measured in an out-of-loop measurement by placing a second photodiode into the main laser path. Below \SI{1}{ms}, we measured that the stability is limited by the photodiode noise.
Between \SI{1}{ms} and \SI{10}{s} the residual relative power fluctuations are $\num{<2e-5}$. Above \SI{10}{s} the stability drifts presumably due to thermal drifts of the photodiode and reaches \num{7e-4} at \SI{1e4}{s}.
\hlbreakable{The influence of these residual power fluctuations on the fluorescence detection noise is shown in figure \ref{fig:fluor_noise}.}

\subsection{Cavity pre-stabilization} \label{sec_cavity}
To resolve Ramsey fringes with a linewidth of around \SI{60}{kHz}, the laser frequency
has to be pre-stabilized to an optical cavity. To enable a frequency offset between the cavity resonance and the laser frequency on atomic resonance, the dual sideband locking technique is used \cite{Thorpe2008}. Using a fibre EOM 
, a Pound-Drever Hall (PDH) modulation frequency of \SI{25}{MHz} and an offset modulation frequency of about \SI{135}{MHz} are applied to the laser light  before coupling $~\SI{100}{\micro \W}$ into an in-vacuum cavity with a measured finesse of 130\,k \cite{Zimmermann2021}. \\
An optical isolator prevents back reflections which otherwise would lead to an etalon between the cavity mirror and the EOM crystal facet which disturbs the lock.
The power spectral density (PSD) of the photodiode noise is about a factor of 10 above the calculated electronic shot noise \( \displaystyle \text{PSD}_V = \frac{2e^2 \eta P}{h \nu R^2}\).
The transfer function of the system was measured yielding a low-pass at \SI{5}{kHz} from the cavity response, a low-pass at \SI{400}{kHz} from the laser response and a delay of \SI{80}{ns} due to the laser response and signal travel time.
Setting appropriate values for the feedback controller, the bandwidth of the system was set to \SI{1.56}{MHz} with a phase margin of \SI{38}{\degree}. 
A slow unlimited integrator to the piezo voltage of the ECDL ensures high gain at DC and prevents mode jumps and railing.  \\
The error signal slope was measured to be \SI{5.5e-5}{V/ Hz} by varying the input offset of the lockbox while in lock and monitoring the light power transmitted through the cavity. 
Measuring the residual error signal and dividing this by the error signal slope yields an in-loop measurement of the residual laser frequency fluctuations. From this, a lower bound for the frequency instability was measured to be $\sigma_y (\tau) > \num{4e-15}/\tau$. A preliminary beat measurement against an optical two-photon rubidium clock via a frequency comb gives an upper bound for the frequency instability of $\sigma_y (\tau) < \num{1.5e-13}/\tau$. 
\hlbreakable{The thermal noise limit is calculated to be \num{6.6e-16}. The commercial cavity design follows Ref. \cite{Nazarova2006} where a sensitivity of \SI{1.5}{kHz/(m/s^2)} in vertical direction and \SI{14}{kHz/(m/s^2)} in horizontal direction is reported. No RAM- or laser power stabilization was employed for the cavity stabilization. \\
We note that due to the potential high feedback bandwidth of the atomic error signal, the cavity performance is not critical for this experiment. For a more compact setup, a monolithic cavity \cite{Zhang2024} with a stability of \num{1e-13} at \SI{10}{ms} would be sufficient.}

\subsection{Frequency-modulation spectroscopy (FMS)} \label{sec_FMS}
\begin{figure}[!t]
\begin{tabularx}{\columnwidth}[b]{Xr}
    \begin{subfigure}[b]{0.5\columnwidth}
    \hspace{-0.6cm}
    \includegraphics[]{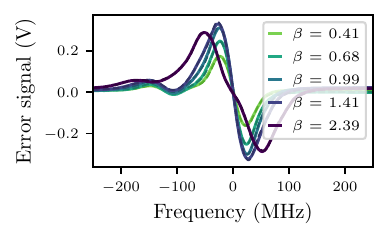}
    \end{subfigure}
 &
    \begin{subfigure}[b]{0.5\columnwidth}
        \hspace{0.1cm}
    \includegraphics[]{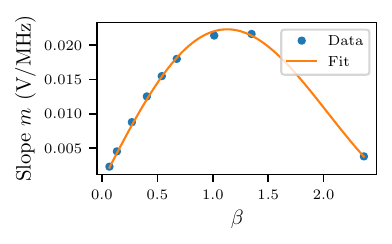} 
    \end{subfigure}     
\end{tabularx} 
    \caption{\textbf{Left:} FMS error signals for varying modulation index $\beta$. Solid lines are experimental data, dashed lines are a theoretical fit with equation \ref{eq_fms}. \\ \textbf{Right:} Error signal slope plotted against modulation index. }
    \label{fig:signal_fms}
\end{figure}

The detection laser at \SI{461}{nm} is locked to the transition frequency via FMS in a single-pass configuration. Modulation is applied by a resonant free space electro-optical modulator 
with resonance frequency of \SI{30.86}{MHz}. The laser beam has a large diameter of \SI{3.5}{mm} ($1/e^2$) to minimize saturation effects. After passing the atomic beam, the photodiode signal 
is split into a DC- and AC-signal. The latter is amplified, filtered and demodulated to generate an error signal, which is fed into a PID controller. The error signal slope was maximized to $m = \SI{59}{mV/MHz}$.\\
A single integrator stage is used for fast feedback to the laser current. The bandwidth of the lock is set to \SI{100}{kHz} as for higher frequencies the detection noise exceeds the free-running laser frequency noise. A higher-order lowpass filter at 5MHz suppresses signals at the modulation frequency. The power spectral density of the detection noise can be modelled by $S_V(f) \approx \SI{4e-10}{V^2/Hz}$ for frequencies higher than \SI{1}{Hz}. The short-term residual frequency noise is limited by detection noise to \( \displaystyle S_\nu(f) \approx S_V(f) / m^2 = \SI{1e5}{Hz^2/Hz} \) which corresponds to an Allan deviation of 
\begin{align}
    \sigma_y(\tau) \approx \num{3e-13}/\sqrt{\tau/1s}\,.
\end{align}
The output of a second stage integrator with a bandwidth of \SI{200}{Hz} is fed back to the piezo voltage to prevent mode jumps of the laser. 
Laser-frequency-dependent residual amplitude modulation originates from etalon effects within the EOM and was minimized by tilting the EOM \cite{Duong2018}. The residual etalon leads to a maximum shift of the laser frequency of \SI{150}{kHz} and a typical drift due to thermal drifts of the crystal of $\sigma_{y}(\tau) \approx \num{2e-11} \, \tau$.  \\
In the following more details are given regarding the optimization of the slope of the error signal. Spectra were obtained for varying modulation index, demodulation phase and laser power, see figure \ref{fig:signal_fms}. A model based on \cite{Silver1992b} was fit to the spectra which is briefly presented in the following.
Both the DC signal ($n=0$) and the error signal ($n=1$) can be modelled by \cite{Silver1992b}
\begin{equation} \label{eq_fms}
\centering
    V = 2 \,G \,I \,\left[\Re(Z) \,\cos(\theta) - \Im(Z) \,\sin(\theta)\right]
\end{equation}
with the laser intensity $I$, gain $G$ and demodulation phase $\theta$. The complex photodiode signal $Z$ is given by
\begin{equation}
\label{eq:freq_mod}
\begin{split}
    Z = &\sum_l J_l(\beta) \,J_{l-n}^*(\beta) \,\times \\
    &\e^{-\frac{1}{2} \alpha(\omega_0 + l \omega_m)} \hspace{2mm} \e^{-\frac{1}{2} \alpha(\omega_0 + (l-n) \omega_m)} \,\times \\
    &\e^{-\i \phi(\omega_0 + l \omega_m)} \hspace{2mm} \e^{\i \phi(\omega_0 + (l-n) \omega_n)}
\end{split}
\end{equation}
with modulation index $\beta$, modulation frequency $\omega_m$ and demodulation frequency $n \omega_m$. The absorption coefficient $\alpha$ is modelled by a sum of six Voigt peaks with frequency offset and relative amplitudes given by the isotope shifts and abundances of the four stable isotopes \cite{Mauger2007}. 
The Doppler broadening is calculated to be $\Delta \nu_\text{D, FWHM} < \SI{10}{MHz}$ due to the atomic beam aperture. This is lower than the natural linewidth and thus the Gaussian contribution to the Voigt profiles was neglected. The dispersion coefficient $\phi$ is calculated from the Kramers-Kronig relation \cite[p.25]{Zinth2013}. \\
Parameters for a maximum error signal slope-to-noise ratio (S/N) were found to be 
\begin{align}
    \theta = \SI{1.014}{\pi} ,\hspace{1cm} \beta = 1.14, \hspace{1cm} I = \SI{7.9}{mW}. 
\end{align}
Due to laser power constraints, the laser power was subsequently set to $I=\SI{3}{mW}$. The error slope was then found to be $\Delta S / \Delta \nu$ = \SI{59}{mV/MHz}. \\

\section{Theoretical model for Ramsey-Bordé signal} \label{app_RB}
The theoretical basis was laid by Bordé \cite{Borde1984a}. We adopt the description of Strathearn \cite{Strathearn2022} which treat the interaction as two consecutive Ramsey interferometers. The excitation probability is parameterized as a combination of a Doppler-free background signal $b$ and a contrast function $c$, which can be shown to be equivalent to Bordé's result. 
\begin{align} \label{eq_fit}
    P_e = b\left(1+c\right)
\end{align}
Starting with this we model the fluorescence signal as a function of $P_e(v_t, v_l)$ with the transversal and longitudinal velocity of the atom. Afterwards, this function is separated into $v_t$ and $v_l$ to reduce computational cost. Lastly we model the background function $b$ and the contrast function $c$ with as few free parameters as possible. \\
The fluorescence signal is proportional to the probability of atoms in the ground state $P^d_g = 1-P^d_e$ at the detection zone. To account for decay and potential slight misalignment between the detection and clock laser we set $P^d_e = k \, P_e$.
Then the fluorescence signal can be calculated as as
\begin{align}
    F &= V \int_{-\infty}^{\infty} dv_l  \int_{-\infty}^{\infty} dv_t n_\gamma(v_l) N(v_t, v_l) \,(1-k\, P_e(v_l, v_t)) 
\end{align}
with the atom number $N(v_t, v_l)$ with transversal velocity $v_t$ and longitudinal velocity $v_l$ and number of photons per atom on the photodetector $n_\gamma (v)$ and voltage per photon $V$. Inserting $P_e$ leads to 
\begin{align}
    F &= Vn_\gamma \left(1 - \int_{-\infty}^{\infty} dv_l  \int_{-\infty}^{\infty} dv_t N(v_t, v_l) \,k \, b\,(1+c) \right) \,.
\end{align}

\subsection{Separating transversal and longitudinal velocity}
Both the background function $b$ as well as the contrast function $c$ depend on the longitudinal and the transversal velocity of the atom, leading to an expensive calculation for a thermal atomic ensemble. This limits the usage of the model for a live analysis of experimentally obtained spectra, which is helpful to understand and optimize the spectrum w.r.t to free set parameters such as laser power, magnetic field or laser beam distance. \\
Here we motivate a semi-empirical fit function and show that it describes the theoretical curve well by comparing it to a Monte-Carlo simulation signal based on Bordé's formalism. \\
We approximate the background function to be only depending on the transversal velocity $b(v_t, v_l) \approx b(v_t, v_l = \bar{v_l})$. For the contrast function $c(v_l, v_t)$, only atoms with a transversal velocity less than the interaction broadening contribute. We approximate $c(v_l, v_t) = c(v_l, v_t = 0)$. Then, the fluorescence signal is separated into $v_l$ and $v_t$
\begin{align}
    F &= Vn_\gamma \left( 1- \int_{-\infty}^{\infty} \,N(v_l) \,(1+c(v_l)) \,dv_l \, \int_{-\infty}^{\infty} N(v_t) \,k\,b(v_t) \,dv_t \right) \\
    &= V\,n_\gamma  \left( 1- \left(1+\int_{-\infty}^{\infty} \,N(v_l) \,c(v_l) \,dv_l \right)\, \int_{-\infty}^{\infty} N(v_t) \,k\,b(v_t) \,dv_t \right) \\
    &=: V\,n_\gamma  \left( 1- \left(1+f_c \right)\, k\,f_b \right) \,.
\end{align}

\subsection{Background function $b$}
\begin{figure}
    \centering
    \includegraphics[]{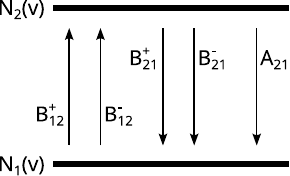}
    \caption{Master equation rate model, see text for details.}
    \label{fig:rate_model}
\end{figure}

The background $b$ models the incoherent envelope of the signal. So neglecting coherent processes we have $b = P_e = 1-P_g = 1 - N_1(v)/N(v)$. Then 
\begin{align}
    f_b = \int_{-\infty}^{\infty} N(v_t) \,b(v_t) \,dv_t = 1 - \int_{-\infty}^{\infty} N_1(v_t)
\end{align}
We model this with a master equation approach as is commonly done for saturation spectroscopy \cite{Foot2005}. For simplicity, we assume two overlapping counter-propagating laser beams (see figure \ref{fig:rate_model}). \footnote{In principle the approach can be extended to separated light fields by using a master equation for every interaction and have the rate of incoming atoms into the ground and excited state scale with the excitation probability of the previous interaction. Then the decay between the interactions is readily implemented which leads to a decrease in the lamb dip. }

\noindent
In a steady state, we have 
\begin{align}
    N_1(v_t) \, (B^+ + B^-) &= N_2(v_t) \,(B^+ + B^- + A)
\end{align}
with light-driven state transfer $B^\pm$ and decay $A$ which includes spontaneous emission and diffusion, see figure \ref{fig:rate_model} Using $N = N_1 + N_2$ as well as introducing $B^\pm / A = s L(\nu \mp v_t/\lambda)$ this leads to 
\begin{align}
    N_1(v_t)  &=\frac{N(v_t)}{2} \left( 1+ \frac{1}{2sL(\nu-v_t/\lambda) + 2sL(\nu+v_t/\lambda) +1} \right) \,.
\end{align}

\noindent
We assume an inhomogeneous broadening due to the Doppler effect and homogeneous broadening due to transit time broadening and power broadening. Collisional broadening is negligible. The transversal velocity is assumed to be a Lorentzian distribution which is a decent approximation to the distribution that was calculated in Appendix \ref{app_oven}. Empirically we also find a better agreement to the experimental data with a Lorentzian compared to a Gaussian as used in \cite{Strathearn2022}.
The homogeneous broadening results from saturation broadening of the natural linewidth and interaction time broadening. The transit time broadening for a Gaussian beam with beam radius $\omega$ and atomic velocity $v$ is a Gaussian with FWHM given by \cite[p.86]{Demtroeder2008b}
\begin{align} \label{equation_tof}
    \delta \nu_{tt} = \sqrt{8 \ln(2)}/\pi \frac{v}{d} \approx 0.75 \frac{v}{d_{1/e^2}}\,.
\end{align}
For simplification, the resulting homogeneous Voigt line was reduced to either a Lorentzian or Gaussian lineshape. Empirically we find a better agreement to the data with a Lorentzian implying the saturation broadening of the Lorentzian natural linewidth is the dominant factor. \\
To conclude, the master equation approach for the background neglects coherent and strong-field effects, implements decay phenomenologically and simplifies the velocity distribution. However, the presented approach can reproduce theoretical and experimental results quite well for a large range of laser power. For a more comprehensive treatment see \cite{Ishikawa1994}.

\subsection{Contrast function $c$}
\begin{figure}
    \centering
    \includegraphics[]{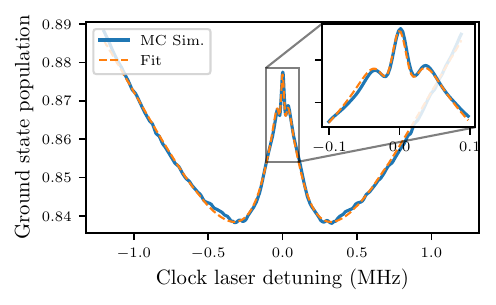} %
    \caption{Comparison of the presented model with a Monte-Carlo simulation based on \cite{Borde1984a}. See text for details.}
    \label{fig_RBI_theory}
\end{figure}

 The contrast function of one recoil port for a single atom with $\pi/2$ pulse area and no decay would be given by 
\begin{align}
    c_{1}^\pm(\nu) = \frac{\eta}{4} \cos(\Phi^\pm)
\end{align}
with a contrast quality factor $0<\eta<1$ and fringe phase 
\begin{align}
    \Phi^\pm = 4\pi \, \frac{d}{v} \left(\Delta \nu \pm \delta\nu_\text{recoil} + \frac{1}{2} \nu_0 \frac{v^2}{c^2}\right) + \phi \,.
\end{align}
The phase $\phi$ is the sum of the optical phases transferred to the atomic wavefunction at each interaction $\phi = \phi_4-\phi_3+\phi_2-\phi_1$. \\
For the overall contrast signal one has to average over the longitudinal velocity distribution. The velocity affects time during the free propagation as well as the duration of the light-atom interaction in the atom interferometer and the detection. It therefore affects the fringe width, the received pulse area, the decay during the interferometer as well as the amount of photons per atom. The latter can be included in the velocity distribution by using the standard Maxwell-Boltzmann distribution instead of the modified distribution for an atomic beam. \\
Averaging over the axial velocity distribution $P_{v_L}(v_L)$ and including decay as well as a velocity-dependent pulse area yields
\begin{align}
    c(\nu) = \frac{\eta}{4} \int_{0}^{\infty} dv P_{v_L}(v_L) \, \left(\cos(\Phi^+) + \cos(\Phi^-)\right) \, \sin^4\left(A \frac{v_m}{v}\right) \,\exp\left(-\frac{D}{v} \Gamma\right) \,.
\end{align}
No normalization for the pulse area or decay term is done, so $\eta$ can now be $>1$. 
The exponential term weights the velocity contribution to the decay. The sinusoidal term weights the velocity contribution to the pulse area A experienced by an atom with mean velocity. The influence of detuning on the pulse area is neglected here. This leads to artefacts for high pulse areas $A \gtrsim \pi$ where several longitudinal velocity classes contribute significantly to the fringe signal.
The integral is approximated by 300 equally spaced velocities between 1m/s and $5\,\bar{v}$. Empirically a frequency offset between the background and the fringes is introduced. This is likely due to a slight beam-pointing misalignment. \\
We find good agreement with the experimental data. We further compared the presented approach to a Monte-Carlo Simulation based on the formulas in \cite{Borde1984a}. For this, we fixed the following values: $V = N_\gamma= 1$, $d = \SI{4}{mm}$, $D = \SI{9}{mm}$, $\phi = \pi$. A good agreement is found as seen in figure \ref{fig_RBI_theory}.

\section{Atomic beam velocity distribution} \label{app_oven}
Here we present a model to predict the atomic beam exiting the oven, which can be compared to experimentally obtained data. The atomic flux $d^3\dot{N}(\theta, v)$ into a solid angle $d^2\Omega$ can be described by \cite{Beijerinck1975}
\begin{align}
    d^3\dot{N}(\theta, v) = I(0) \, f(\theta) F(v) \Gamma_{\theta}(v) \, d^2\Omega\, dv
\end{align}
with the axial intensity $I(0)$, angular distribution profile $f(\theta)$, velocity distribution function $F(v)$ and deformation function $\Gamma_{\theta}(v)$. The angle $\theta$ is measured to the principal atomic axis. The axial intensity of a reservoir with vapour number density $n$, average velocity $\bar{v}$ and orifice area $A$ is given by $I(0) = \frac{n A \bar{v}}{4 \pi}$. The velocity distribution $F(v)$ of a thermal atomic beam is given by 
\begin{align}
            &F(v) = \frac{2}{\alpha} \left( \frac{v}{\alpha} \right)^3 \exp\left[ - \left( \frac{v}{\alpha} \right)^2 \right] \,.
\end{align}
The additional $v$ term compared to the standard Maxwell-Boltzmann distribution stems from faster atoms having a higher probability of exiting the orifice in a given time frame. However, the amount of photons emitted per atom within the detection zone scales inversely with the velocity. Thus for comparison to fluorescence spectra, a standard Maxwell-Boltzmann distribution is used. If the temperature of the capillaries is similar to the reservoir temperature, the collimation process does not alter the absolute velocity distribution significantly, thus usually $\Gamma_\theta(v) = 1$ is assumed. \\
Commonly it is approximated that the axial intensity is unaffected by the capillary array. In the transparent regime where the mean free path of inter-atomic collision $\lambda$ is larger than the capillary length and thus inter-atomic collisions can be neglected, the angular distribution profile is given by \cite{Beijerinck1975}
\begin{align}
            f(\theta) =  \tfrac{2}{\pi}\cos(\theta) \left((1-\tfrac{1}{2}W)R(\rho) + \tfrac{2}{3}(1-W) [1-(1-\rho^2)^{3/2}] \rho^{-1}\right) +\tfrac{1}{2} W \cos(\theta) 
\end{align}
for angles $|\theta| < \arctan(2a/L)$ and 
\begin{align}
            f(\theta) =  \frac{1}{\pi}\frac{\cos^2(\theta)}{\sin(\theta)} \frac{8a}{3L} (1-W) + \tfrac{1}{2}W \cos(\theta)
\end{align}
for angles $|\theta| > \arctan(2a/L)$. Here, $\rho = \frac{L}{2a} \tan(\theta)$ and $R(\rho = \arccos(\rho) - \rho(1-\rho^2)^{\tfrac{1}{2}}$.
For the Clausing factor $W$ various approximations exist which can be compared to Monte-Carlo simulations. The most common approximation is $W \approx \frac{4}{3\Gamma + 4}$ with $\Gamma = L/(2a)$. We used the more accurate approximation \cite{Lucas1973a}
\begin{align}
W \approx 2 (2\Gamma +7)/(3\Gamma^2 +18\Gamma+14) \,.
\end{align}
To compare to experimentally observed spectra, we first generate atoms with a Monte-Carlo simulation using $f(\theta)$ and $F(v)$ in cylinder coordinates with equally distributed $f(\phi)$ and starting position $(x_0, y_0)$ within the limits of the oven orifice. After switching to Cartesian coordinates, atoms are removed that either hit an aperture or are more than $1/e^2$ diameter away from the laser beam centre. The remaining atoms are binned according to velocity in laser beam direction. 
Lastly, a convolution with a power-broadened Lorentzian gives the expected spectrum which can be compared. Slightly counter-intuitively, a laser beam displacement in the y-direction leads to a broadening in the distribution in the x-direction. We assume this mechanism to be the cause of the broadening seen in figure \ref{fig_oven_complete}\,(c). \\
In the case of the RBI setup, the angle given by the apertures is much smaller than the capillary aspect ratio, thus the profile is dominated by the apertures. Then, a decent approximation for the velocity distribution in laser beam direction is a Lorentzian, which is slightly wider in the centre and at the wings.   

\end{appendices}

\bmhead{Acknowledgements}
We would like to thank Matthias Schoch for his help with the fluorescence detector, Levi Wihan for the wavefront measurement of the retroreflector as well as Klaus Döringshoff, Julien Kluge, Moritz Eisebitt and Daniel Kohl for their help with the beat measurement between the cavity stabilized \SI{689}{nm} laser and the two-photon rubidium clock. We would like to thank the OPUS and QUASENS collaboration for fruitful discussions.

\bmhead{Author contributions}
OF, CZ, MJ and VS built the experimental setup. 
OF wrote the first draft of the manuscript. IT, AM, CZ and MK reviewed and edited the manuscript. MK acquired funding and supervised the project. All authors have read and agreed to the submitted version of the manuscript.

\bmhead{Funding} 
This work is supported by the German Space Agency (DLR) with funds provided by the Federal Ministry of Economic Affairs and Climate Action (BMWK) under grant numbers 50WM1852, 50WM2250B and by the Federal Ministry of Education and Research within the program quantum technologies - from basic research to market under grant number 13N15725.

\section*{Declarations}

\bmhead{Financial interests}
The authors have no relevant financial or non-financial interests to disclose.

\bmhead{Ethics approval and consent to participate}
Not applicable

\bmhead{Consent for publication}
Not applicable

\bmhead{Competing interests}
The authors declare no competing interests.


\bibliography{OPUS-For_RBI_Paper}

\end{document}